\documentclass[onefignum,onetabnum]{siamart171218}

\def\b#1{{\bf #1}}   

\usepackage{lipsum}
\usepackage{amsfonts}
\usepackage{graphicx}
\usepackage{epstopdf}
\usepackage{algorithmic}
\usepackage{comment}
\ifpdf
  \DeclareGraphicsExtensions{.eps,.pdf,.png,.jpg}
\else
  \DeclareGraphicsExtensions{.eps}
\fi

\newsiamremark{remark}{Remark}
\newsiamremark{hypothesis}{Hypothesis}
\crefname{hypothesis}{Hypothesis}{Hypotheses}
\newsiamthm{claim}{Claim}

\headers{Weighted Rank Aggregation}{T. Perini, A. Langville, and xCures}

\title{Weight Set Decomposition for Weighted Rank Aggregation: An Interpretable and Visual Decision Support Tool
}

\author{Tyler Perini\thanks{Department of Computational and Applied Mathematics, Rice University, Houston, TX 
  (\email{tyler.perini@rice.edu}, \url{http://www.tylerperini.com}).}
\and Amy N. Langville\thanks{Department of Mathematics, College of Charleston, Charleston, SC 
  (\email{langvillea@cofc.edu}).}
  \and Glenn Kramer\thanks{Chief Technology Officer, xCures, Inc. 
  (\email{gkramer@xcures.com}).}
\and Jeff Shrager\thanks{Director of Research, xCures, Inc. 
  (\email{jshrager@xcures.com}).}
  \and Mark Shapiro\thanks{Chief Operating Officer, xCures, Inc.
  (\email{mshapiro@xcures.com}).}
  }

\usepackage{amsopn}

\makeatletter
\newcommand*{\addFileDependency}[1]{
  \typeout{(#1)}
  \@addtofilelist{#1}
  \IfFileExists{#1}{}{\typeout{No file #1.}}
}
\makeatother

\newcommand*{\myexternaldocument}[1]{%
    \externaldocument{#1}%
    \addFileDependency{#1.tex}%
    \addFileDependency{#1.aux}%
}

\usepackage{graphicx}

\newcommand{\R}{\mathbb{R}}

\newcommand{\Z}{\mathbb{Z}}

\ifpdf
\hypersetup{
  pdftitle={Weighted Rank Aggregation},
  pdfauthor={T. Perini, A. N. Langville, and G. Kramer, J. Shrager, M. Shapiro, and A. Wasserman}
}
\fi

\myexternaldocument{ex_supplement}

\begin{document}

\maketitle

\begin{abstract}
The problem of interpreting or aggregating multiple rankings is common to many real-world applications. Perhaps the simplest and most common approach is a weighted rank aggregation, wherein a (convex) weight is applied to each input ranking and then ordered. This paper describes a new tool for visualizing and displaying ranking information for the weighted rank aggregation method. Traditionally, the aim of rank aggregation is to summarize the information from the input rankings and provide one final ranking that hopefully represents a more accurate or truthful result than any one input ranking. While such an aggregated ranking is, and clearly has been, useful to many applications, it also obscures information. In this paper, we show the wealth of information that is available for the weighted rank aggregation problem due to its structure. We apply weight set decomposition to the set of convex multipliers, study the properties useful for understanding this decomposition, and visualize the indifference regions. This methodology reveals information--that is otherwise collapsed by the aggregated ranking--into a useful, interpretable, and intuitive decision support tool. Included are multiple illustrative examples, along with heuristic and exact algorithms for computing the weight set decomposition.  
\end{abstract}



%

\section{Introduction}
\label{section:introduction}

Ranking is a common task in many applied fields including machine learning, search, sports analytics/scheduling, economic forecasting, and health care. Ranking is foundational to hundreds of algorithms designed by companies such as Netflix, Amazon, and Google. Google's algorithm for ranking the webpages relevant to a user's search query combines hundreds of ranking measures. The combination of such measures is typically done with an \emph{aggregation method}. One of the simplest, and, therefore, perhaps most common, aggregation methods is some version of a weighted average, weighted sum, linear combination, or convex combination. Consider the following real-world applications.

\begin{itemize}
\item 
Prior to the Bowl Championship Series, U.S. college football teams were selected for post-season bowl games from a ranking that aggregated several rankings from coaches polls of rankings created by human experts as well as the rankings created by computer models, including the Massey and Colley models. The aggregated ranking $\b r^a=\lambda_1 \b r^1 + \lambda_2 \b r^2 + \lambda_3 \b r^3 + \lambda_4 \b r^4 + \lambda_5 \b r^5 + \lambda_6 \b r^6$, where $\lambda_i$ are the respective weights of the six rankings. Ratings can also be aggregated as discussed in Section \ref{section:practicalnotes}. 

\item 
The U.S. News \& World Report creates its annual ranking  $\b r^a$ of American colleges by aggregating 17 features $\b r^1$,\ldots, $\b r^{17}$, such as graduation rate, acceptance rate, and faculty-to-student ratio, with a weighted average. For example, the graduation rate  contributes 8\% while the faculty-to-student ratio contributes 1\%.

\item 
A cancer research company creates a ranking of treatments tailored to individual patients by aggregating rankings from several sources of data. For exposition, we describe three sources of data: biomarker data, data from expert oncologists, and clinical trials data. Each type of data creates a ranking of treatments for the particular patient. A weighted combination aggregates these three rankings into one final aggregated ranking $\b r^a =\lambda_1 \b r^1 + \lambda_2 \b r^2 + \lambda_3 \b r^3$, where $\lambda_i$ are the weights given to each of the three rankings.

\end{itemize}

The aim of rank aggregation is to \emph{summarize} the information from the input rankings and provide \emph{one} final ranking, which should somehow represent a more accurate or truthful result. While such an aggregated ranking is, and clearly has been, useful to many applications, it also \emph{obscures information}. Furthermore, after the aggregated rank is produced, the summarization of information is often under-utilized. In this paper, we show the wealth of information that is available for the weighted rank aggregation problem due to its structure. In other words, we expand the information collapsed by the aggregated ranking with the proposed weight set decomposition and a powerful visualization tool. 

\subsection{Related Work}
\label{section:relatedwork}

This work builds on three related topics. First is a 2007 patent issued to Kramer \cite{Kramer:patent} that describes a method for re-ranking search results based on weights of three factors. The idea and tool described in the patent can be applied to diverse applications such as web search, product search, database search, financial planning, and survey results. The patent also describes the graphical user interface for an interactive tool that allows users to move sliders to set the weights of the factors. Other visualization attempts for the weighted aggregation problem appear in \cite{Schimek2015,Kidwell2008}. Our method, described in Section \ref{section:runningxCures}, is much more intuitive and powerful.

Second, there is weight space decomposition methods from multiobjective optimization (MOO) literature. MOO problems have a set of solutions, called Pareto efficient solutions or the Pareto frontier, which characterize the optimal trade offs between conflicting objectives \cite{ehrgott2005book}.  
A popular utility function used to reduce a MOO problem to a single-objective problem is a \emph{weighted preference model}, e.g., weighted sum of the objectives, since the weights intuitively represent the preferences of the decision maker. For three objectives, the set of possible weights is represented by a triangular set, and then this weight set may be decomposed into regions which map every weight to its resulting Pareto efficient solution. The two most common weighted preference models are weighted sum (1-norm) \cite{benson2000weightset, przybylski2010wsum, alves2016graphical} and weighted Tchebychev ($\infty$-norm) \cite{karakaya2021evaluating, perini2021thesis}. The former has existed in the MOO literature for over a decade and has been well-studied. 

Third, there is work on the rankability of ranking data. Anderson et al. defined the rankability of pairwise ranking data as its ability to produced a meaningful ranking of its items \cite{SIMODS}. In subsequent papers, they link rankability to the cardinality of the set $P$ of multiple optimal rankings \cite{FODS,Anderson2021fairness,Cameron2020}. Thus, this paper's set $A$ of aggregated rankings is related to their set $P$. Further, the rank colormap of Section \ref{section:rankcolormap} is an immediate visual representation of the rankability of the weighted rank aggregation problem. 

\section{Preliminaries}

\subsection{Notation}

Let there be $0<n<\infty$ items to rank, and let $v^j_i\in\R$ be the given value of the $i$th item in the $j$th scoring vector for $i=1..n$ and $j=1,2,3$. Denote $V^j:=\{v^j_i\}_{i=1..n}$ as the  \emph{$j$th input}. Note that we do not assume values are integral, which is the case that $V^j$ is a ranked order. Therefore our method generalizes to whether the input values are \emph{rankings} or \emph{ratings}. Let $\sigma: \R^n \rightarrow \{1, \dots, n\}^n$ be such that $\sigma(v^j_1, \dots, v^j_n)$ return an \emph{acceptable} ranked position of the scores such that if $v^j_i < v^j_{i'}$, then $\sigma_i(V^j) < \sigma_{i'}(V^j)$ (and if $v^j_i \leq v^j_{i'}$, then $\sigma_i(V^j) \leq \sigma_{i'}(V^j)$) for all $i\in\{1..n\}$; therefore, $\sigma_i(V^j)$ represents the \emph{ranked position} of item $i$ given input $j$. Note that without tied scores in any input, then there is a distinct set of ranks; in the case of tied scores, e.g. $v^j_{i}=v^j_{i'}$, then $\sigma$ simply returns one of the acceptable ranked orders. Therefore, $\sigma$ is a function which returns one discrete vector (potentially from a set of possible vectors). 

Let $\Lambda = \{\lambda\in\R^3_+: \lambda_1 + \lambda_2 + \lambda_3 = 1\}$ be the set of convex weights which we call the \emph{weight set}. For any $\lambda\in\Lambda$, denote $v^\lambda := \sum_{i=1}^3 \lambda_i v_i$ as the weighted average of input scores with respect to $\lambda$. Note then that $\sigma(v^\lambda)$ returns a vector of ranked positions for this weighted scoring vector. For a fixed vector of ranked positions, $\Bar{\sigma}$, we define its corresponding \emph{indifference region} by 
$$\Lambda(\Bar{\sigma}) := \{\lambda \in \Lambda: \Bar{\sigma} = \sigma(v^\lambda) \}. $$

\subsection{Running Health Care Example with $j=3$ rankings}
\label{section:runningxCures}

\begin{figure}[h!]
\centering
\includegraphics[height=5cm]{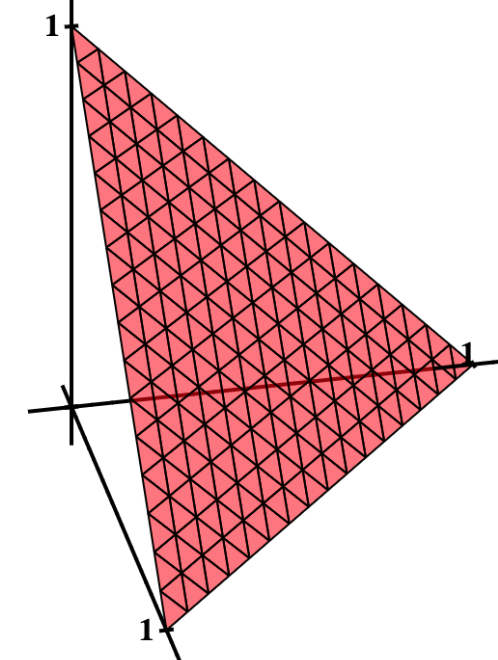} \includegraphics[height=5cm]{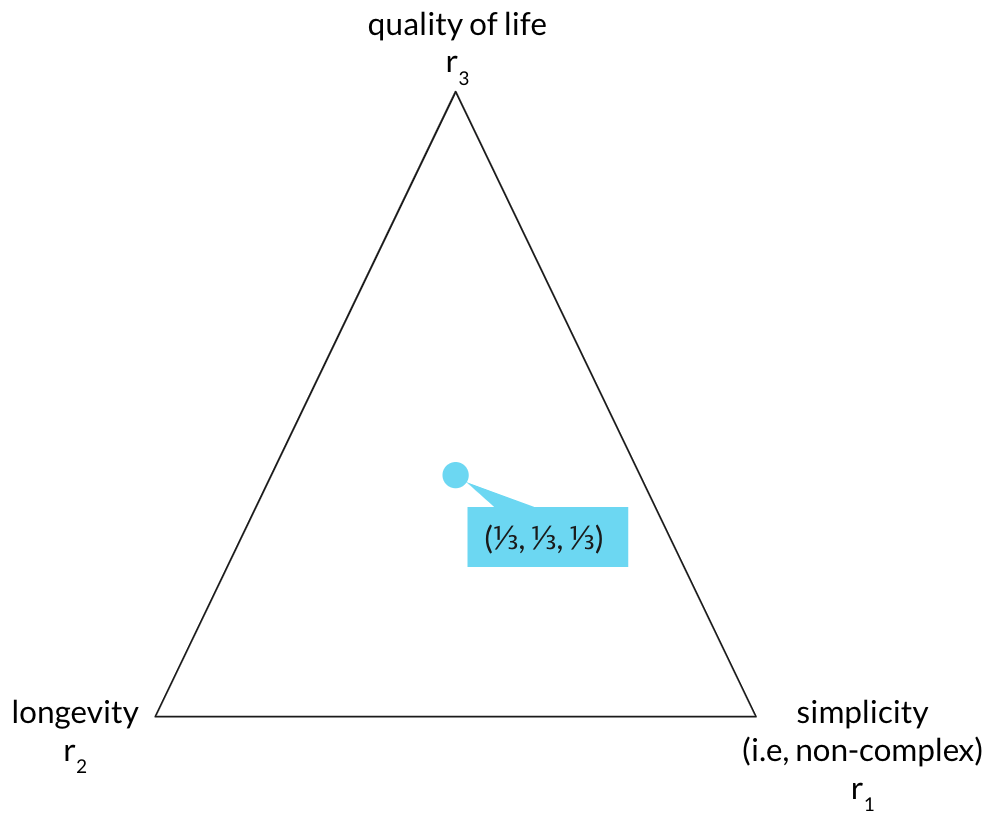}
\caption{The set of convex weights $\Lambda$ creates a triangle in $\Re^3$ (left) that can be visualized in $\Re^2$ (right). When the weights are equally weighted, i.e., $\lambda = (\frac{1}{3}, \frac{1}{3}, \frac{1}{3})$, the aggregated ranking is equidistant from the three corners, and hence, the three input rankings.
}
\label{fig:LambdaTriangle}
\end{figure}

In order to tailor a treatment program to an individual cancer patient, the oncologist asks the patient to give weights (e.g., by moving slider bars or assigning percentages) to indicate the relative importance of three different criteria: complexity, effectiveness, and quality of life. Suppose Patient A, Anne, scores these three criteria as 6, 4, 2, respectively. These can be normalized to create the convex combination of weights $(\frac{1}{2}, \frac{1}{3}, \frac{1}{6})$. We call the set of convex weights $\Lambda$ the \emph{weight set}, where $\Lambda := \{\lambda\in\R^3_+: \lambda_1 + \lambda_2 + \lambda_3 = 1\}$. We can visualize $\Lambda$, which is the intersection of the $\Re^3$ plane $\lambda_1 + \lambda_2 + \lambda_3 = 1$ and the nonnegative octant. See Figure \ref{fig:LambdaTriangle}.

A cancer research company uses machine learning to create 
the three treatment rankings $\b r^1$, $\b r^2$, $\b r^3$ (shown below), tailored to Anne's individual patient data, shown below for the three criteria (complexity, effectiveness, and quality of life, respectively). Treatment $i$ is denoted $T_i$, e.g., Treatment $T_1$ is Temozolmide.\footnote{These five cancer treatment drugs were chosen for explanatory purposes only.} Notice that the rankings differ. For example, the \#1 treatment for Anne with respect to quality of life is $T_2$ Pembrolizumab, yet for both complexity and effectiveness the \#1 treatment is $T_1$ Temozolmide, which appears last for quality of life. While the rankings $\b r^1$, $\b r^2$, $\b r^3$ below are complete lists, our methods and tool also handle incomplete lists. See Section \ref{section:incompletelists}.

$$
\b r^1 = \bordermatrix{& \hbox{\textbf{complexity}} \cr
1^{st} & \hbox{$T_1$ Temozolomide} \cr
 2^{nd}& \hbox{$T_2$ Pembrolizumab} \cr 
 3^{rd} & \hbox{$T_3$ Gliovac}\cr 
 4^{th}  & \hbox{$T_4$ Bevacizumab} \cr 
 5^{th}   & \hbox{$T_5$ Adavosertib} \cr},
 $$
 
 $$
 \b r^2 = \bordermatrix{& \hbox{\textbf{effectiveness}} \cr
1^{st} & \hbox{$T_1$ Temozolomide} \cr
2^{nd} & \hbox{$T_3$ Gliovac} \cr 
3^{rd}  & \hbox{$T_2$ Pembrolizumab} \cr 
 4^{th}   & \hbox{$T_4$ Bevacizumab} \cr 
 5^{th}   & \hbox{$T_5$ Adavosertib} \cr}, 
 $$
 
 $$
 \b r^3 = \bordermatrix{& \hbox{\textbf{quality of life}} \cr
1^{st} &  \hbox{$T_2$ Pembrolizumab}\cr 
2^{nd} & \hbox{$T_3$ Gliovac} \cr 
3^{rd}  &  \hbox{$T_4$ Bevacizumab} \cr
 4^{th}   & \hbox{$T_5$ Adavosertib} \cr 
 5^{th}   & \hbox{$T_1$ Temozolomide} \cr}. 
$$

\begin{figure}
\centering
\includegraphics[height=5cm]{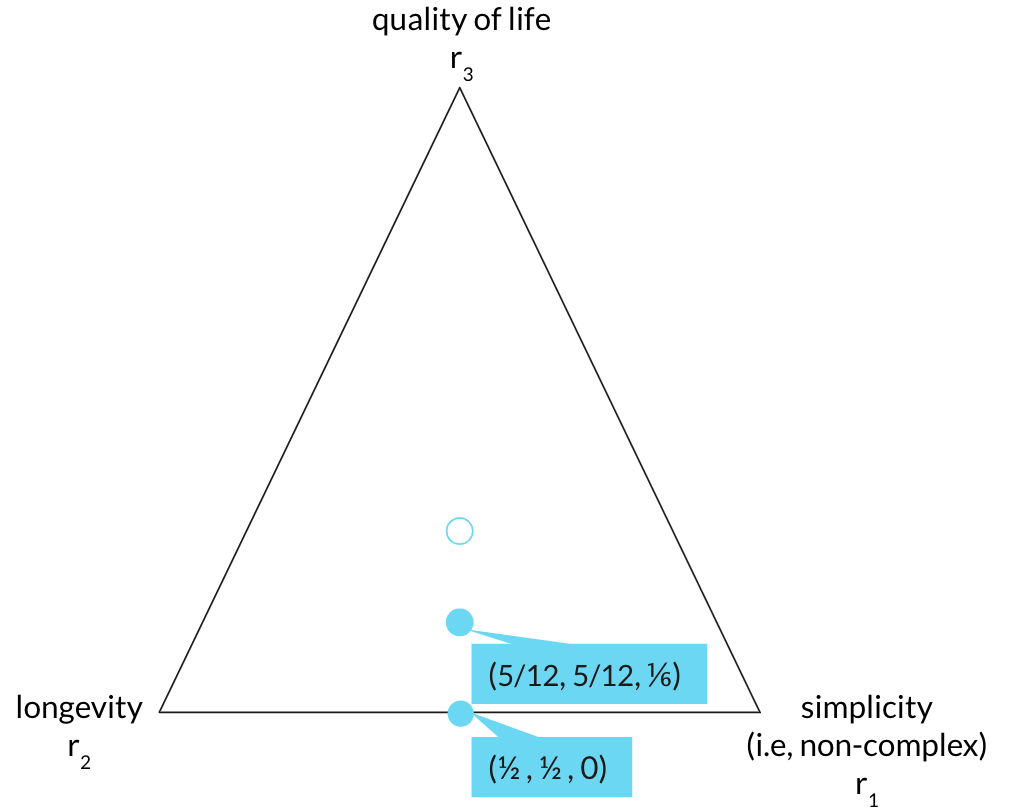}\hskip.2in \includegraphics[height=6cm]{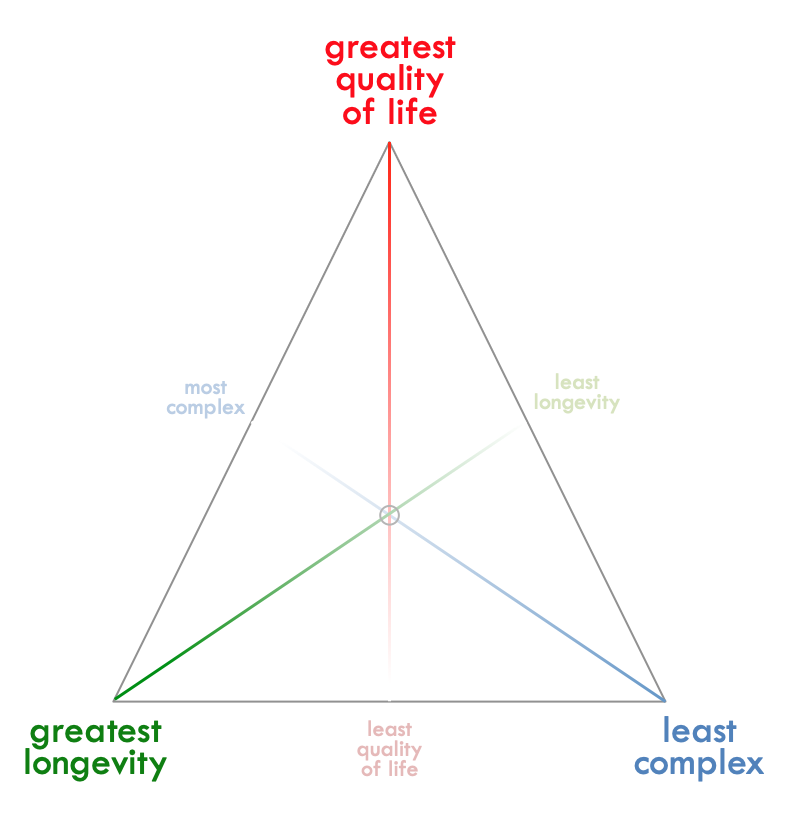}
\caption{The triangle on the left shows the weight set $\Lambda$ for Patient A, Anne, visualized in $\Re^2$. Two sets of weights are shown. The weights $(\frac{1}{2}, \frac{1}{2}, 0)$ is midway along the boundary between $\b r^1$ and $\b r^2$, and so it compromises between only two criteria (longevity and simplicity).  The weight $(\frac{5}{12}, \frac{5}{12}, \frac{1}{6})$ is in the interior of the triangle, moved slightly toward the third criterion of cost.  The triangle on the right shows colored axes that indicate the relative weight of a goal and guide users in movement toward or away from a particular goal.
}
\label{fig:AnneLambdaTriangle}
\end{figure}

We can visualize and contextualize Anne's preferences using $\Lambda$ in 2D, as shown in Figure \ref{fig:AnneLambdaTriangle} (left). Each corner represents one of the $j=3$ (input) ranking vectors with its corresponding criteria labeled: 
\begin{itemize}
    \item The simplicity (i.e., non-complexity) ranking $\b r^1$ occurs when $\lambda = (1, 0, 0)$, which is the right corner of the triangle. 
    \item The longevity ranking $\b r^2$ occurs when $\lambda = (0, 1, 0)$, which is the left corner of the triangle. 
    \item The quality of life ranking $\b r^3$ occurs when $\lambda = (0, 0, 1)$, which is the top corner of the triangle.
\end{itemize}

The convex hull of the three extreme points (corners) defines the triangle of infinitely many (convex combination) weights from which Anne and her doctor might choose. For example, her doctor may first recommend equal weights, as shown in Figure \ref{fig:LambdaTriangle}, and computed as:  
$$\b r^a = \frac{1}{3}\b r^1 + \frac{1}{3}\b r^2 + \frac{1}{3}\b r^3 = (1.3, 2.6, 3.0, 4.3, 3.6) \rightarrow (1, 2, 3, 5, 4),$$
which results in the ranking of $[1 \;\;  2\;\;  3 \;\; 5\;\;  4]^T$ and hence ranks treatments from first to last place as $T_1$ Temozolomide, $T_2$ Pembrolizumab, $T_3$ Gliovac, $T_5$ Adavosertib, $T_4$ Bevacizumab.
After discussion, Anne and her family decide to update her preferred weights to $(\frac{1}{2}, \frac{1}{2}, 0)$.  For the new weights, 
$$\b r' = \frac{1}{2}\b r^1 + \frac{1}{2}\b r^2 + 0\b r^3 = (1.0, 2.5, 2.5, 4.0, 5.0) \rightarrow (1, 2, 3, 4, 5) \ \text{or}\ (1, 3, 2, 4, 5),$$
which induces a tie between the second and third item. 
A compromise between those two weights vectors could be computed by taking the average of the weights, creating $(\frac{5}{12},  \frac{5}{12}, \frac{1}{6})$:
$$\b r'' = \frac{5}{12}\b r^1 + \frac{5}{12}\b r^2 + \frac{1}{6}\b r^3 = (1.16, 2.58, 2.75, 4.16, 4.33) \rightarrow (1, 2, 3, 4, 5).$$
Observe that these three weights correspond in $\Lambda$ to either \emph{moving towards or away from} the extreme point representing quality of life, where the latter moves towards the lower boundary between extreme points for longevity and simplicity (where the weight for quality of life is zero). The three axes in Figure \ref{fig:AnneLambdaTriangle} (right) represent the relative weight in each of the three criteria, and they provide users an intuition for choosing and updating preferences within the weight set.  

Note that each $\lambda$ vector maps to an aggregated ranking $\b r^a$. Yet even though there are infinitely many weight vectors in $\Lambda$, there is a finite number of possible aggregated rankings $\b r^a$ that result. For given input of $\lambda\in \Lambda$, $\b r^1$, $\b r^2$, and $\b r^3$, how many aggregated rankings can be output? Mathematically, we have an upperbound on the number of aggregated rankings since $n$ items can be rearranged in at most $n!$ ways; however, we rarely expect this bound to be tight. In this paper, our goals are to enumerate and visually display these aggregated rankings in order to glean novel and useful ranking information.

\subsection{Rank Colormap}
\label{section:rankcolormap}

\begin{figure}[h!]
\centering
\includegraphics[height=5cm]{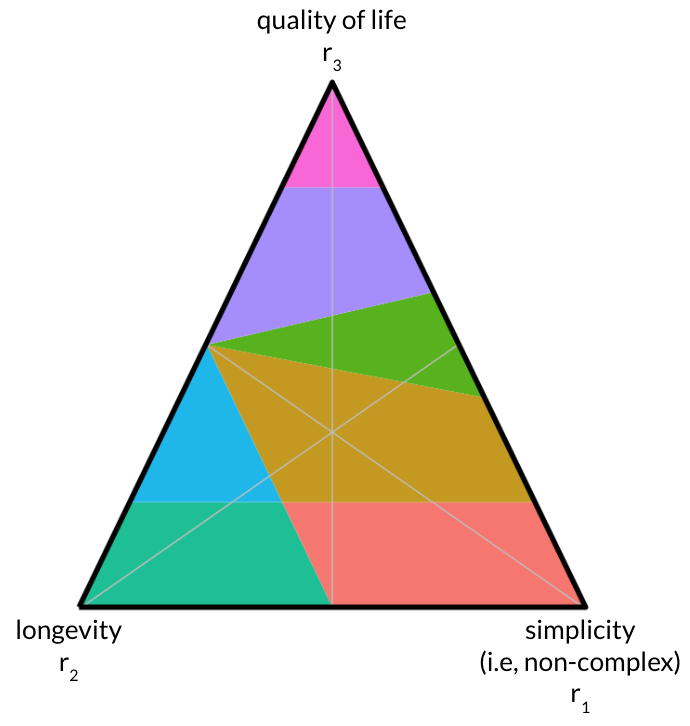} 
\includegraphics[height=5cm]{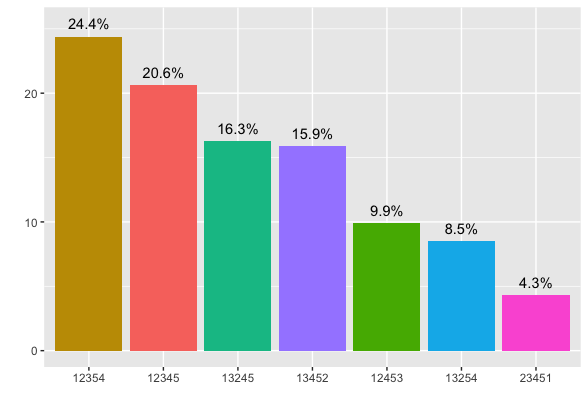} 

\caption{(Left) Rank colormap for Patient A, Anne. (Right) Barchart displaying the percentage of $\Lambda$ on the $y$-axis per region with the corresponding ranking labeled on the $x$-axis. A quick scan of this barchart shows the relative area associated with each ranking.  
}
\label{fig:AnneColormap}
\end{figure}

We define the \emph{rank colormap} as the mapping of each weight in $\Lambda$ to its corresponding aggregated ranking. Figure \ref{fig:AnneColormap} shows Anne's rank colormap. 
Notice that the central weight could be perturbed from $\lambda=(\frac{1}{3}, \frac{1}{3}, \frac{1}{3})$ a bit--say to $(.3, .3, .4)$ or even $(.75, .20, .05)$--and still result in the same ranking of treatments. In fact, any $\Lambda$ in the mustard-colored region (which includes the central weight) produces that same ranking. 
For a given ranking $\b r$, we call the set $\Lambda(\b r) := \{\lambda\in\Lambda: \sum_j \lambda_j r^j = \b r \}$ the \emph{indifference region} for $\b r$ because the aggregated ranking is indifferent to which weight is chosen from this set. 
Observe that Anne's original weight $\lambda=(\frac{1}{2}, \frac{1}{2}, 0)$ is at the intersection of the green and salmon regions, which correlates to a tie between the two corresponding ranks.

The rank colormap summarizes a great deal of ranking information:
\begin{enumerate}
    \item For Anne, there are 7 distinct indifference regions, one for each of the possible aggregated rankings. Let $A$ be the set of possible aggregated rankings. The labels below each bar of the barchart of Fig. \ref{fig:AnneColormap} (right) show that, for Anne's data, $A=\{[1 \; 2 \; 3 \; 5 \; 4], [1 \; 2 \; 3 \; 4 \; 5], [1 \; 3 \; 2 \; 4 \; 5], [1 \; 3 \; 4 \; 5 \; 2], [1 \; 2 \; 4 \; 5 \; 3], [1 \; 3 \; 2 \; 5 \; 4], [2 \; 3 \; 4 \; 5 \; 1]\}$.
    The cardinality of this set, $|A|$ =7, is an indication of the \emph{rankability} of the data \cite{SIMODS,FODS,Cameron2020,McJamesRankability}, a topic we will discussed earlier in Section \ref{section:relatedwork}.  
    
    \item The \emph{area} of an indifference region in the colormap indicates the aggregated ranking's sensitivity to small changes in the weights. The bar chart on the right of Figure \ref{fig:AnneColormap} shows both the number of aggregated rankings as well as the area associated with each aggregated ranking. An indifference region  with small area (e.g., ranking $[2 \; 3 \; 4 \; 5 \; 1]$ at 4.3\%) that is sensitive to these changes is generally less preferred than an indifference region with a large area (e.g., ranking $[1 \; 2 \; 3 \; 5 \; 4]$ at 24.4\%), which is insensitive or \emph{robust}. For Anne, quality of life ranking $\b r^3$ is the most sensitive ranking, since the pink region at the top of the colormap has the smallest area. The longevity ranking $\b r^2$ is slightly more sensitive than the simplicity ranking $\b r^1$.
    
    \item Indifference regions in the colormap clearly indicate \emph{adjacency} between aggregated rankings.  Each neighboring region in the colormap is just one swap away from one another. These adjacencies may be useful in designing \emph{paths of sequential solutions}, e.g., for adjusting a patient's treatment in response to outcomes.

    \item Two $n \times n$ matrices $\b X^*$ and $\b A^*$ are created from the set $A$ of aggregated rankings. The $(i,j)$ element of $\b X^*$ is the number of rankings in $A$ ranking item $i$ above item $j$. For example in the matrix below $\b X^*(1,5) = \frac{6}{7}$ because 6 of the 7 rankings in $A$ have Treatment 1 ranked above Treatment 5. As a result, $\b X^*$ has structure, i.e., $\b X^*(i,j) = 1-\b X^*(j,i)$. Another matrix $\b A$ is created so that the $(i,j)$ element of $\b A^*$ is the percentage of the area of the colormap in $A$ ranking item $i$ above item $j$.
    $$
    \b X^*=\bordermatrix{ & T_1 & T_2 & T_3 & T_4 & T_5 \cr
    T_1 & 0 & \frac{6}{7}& \frac{6}{7}& \frac{6}{7}&  \frac{6}{7} \cr
    T_2 & \frac{1}{7}& 0& \frac{5}{7}& \frac{6}{7}&\frac{6}{7} \cr
    T_3 & \frac{1}{7} & \frac{2}{7} &0 &\frac{6}{7} &\frac{6}{7} \cr
    T_4 &\frac{1}{7} &\frac{1}{7} &\frac{1}{7} &0 &\frac{5}{7} \cr
    T_5 & \frac{1}{7}&\frac{1}{7} &\frac{1}{7} & \frac{2}{7}& 0\cr} \hbox{and }
    \b A^*=\bordermatrix{ & T_1 & T_2 & T_3 & T_4 & T_5 \cr
    T_1 & 0 & .96 & .96 & .96 & .96 \cr
    T_2 & .04 & 0& .59& .84 & .84 \cr
    T_3 & .04 & .41 & 0 & .90 & .90 \cr
    T_4 & .04 & .16 & .10 & 0 & .67 \cr
    T_5 & .04 & .16 & .10 & .33 & 0 \cr}.
    $$

    \item Due to the similarity between convex combination weights and probabilities, the colormap provides a straightforward stochastic interpretation. Suppose one input rank was randomly drawn with uniform probability. Then the colormap gives information about the \emph{expected ranking}. The number above each ranking's bar in the barchart is that ranking's percentage of the triangle. In other words, it is the probability of seeing that ranking. Thus, the expected ranking is the sum of the probabilistically weighted rankings.
    
\end{enumerate}

\begin{figure}[h!]
\centering
\includegraphics[width=0.44\textwidth]{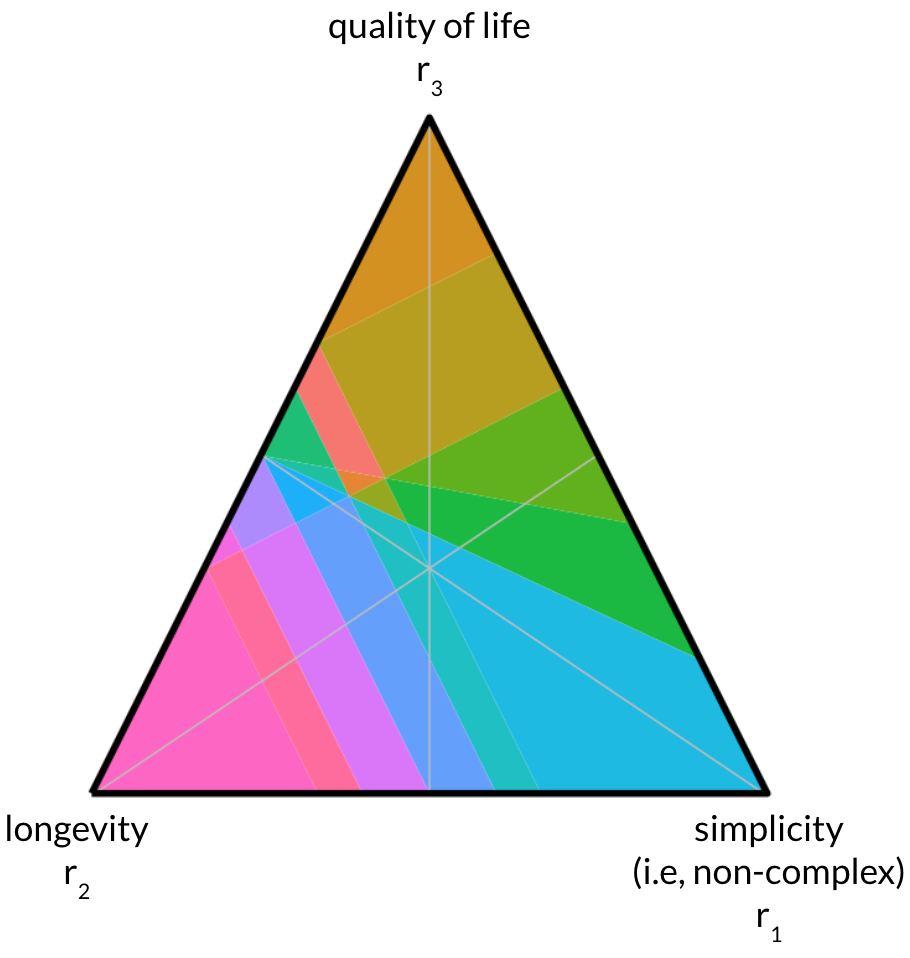} 
\includegraphics[width=0.54\textwidth]{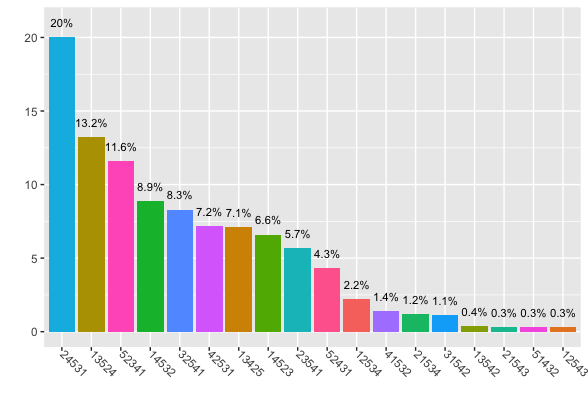} 
\caption{Rank colormap for Patient B, Bob. While Bob's colormap contains $|A|=18$ regions, more than Anne's, several regions, and hence rankings, have insignificant areas.
}
\label{fig:BobColormap}
\end{figure}

Figure \ref{fig:BobColormap} shows the colormap for Patient B, Bob. In Bob's case, there are $|A|=18$ regions that result from aggregation of Bob's three treatment rankings. And yet even with more alternative aggregate rankings, one of the input rankings, $\b r^1$, dominates a large portion of the colormap, as is also clearly shown in the barchart. The barchart shows that several rankings occupy insignificant area in the colormap. These rankings appear in the central left part of the colormap, showing this area is highly sensitive to the weights selected.

\section{Results}

In this section, we construct separating hyperplanes to prove convexity of the indifference regions. 

\subsection{Separating Hyperplanes}
\label{sec:separatinghyperplanes}

Consider distinct items $a$ and $b$ in the input rankings. In the case that all three rankings agree that $r^i_a < r^i_b $ for all $i=1,2,3$ (or  $r^i_a > r^i_b $ for all $i=1,2,3$), then there is no disagreement to be considered. 

So let $a$ and $b$ be such that $r^i_a < r^i_b $ for some $i\in\{1,2,3\}$ and $r^j_a > r^j_b $ for some $j\in\{1,2,3\}$. Then we may partition $\Lambda$ into a subset where the weighted aggregate ranks $a$ (strictly) better than $b$, a subset where the weighted aggregate ranks $b$ (strictly) better than $a$, and a subset where the weighted aggregate has $a$ and $b$ \emph{tied}. We use the final set to define the previous two sets.

Let the \emph{separating hyperplane} be defined by 
$$H(a,b):=\{\lambda\in\Lambda: \lambda^T [r^1_a, r^2_a, r^3_a] = \lambda^T [r^1_b, r^2_b, r^3_b] \}.$$
We begin by demonstrating the equation representing the line containing $H(a,b)$, followed by computing the endpoints of line segment $H(a,b)$.

Let $\delta^i=r^i_a-r^i_b$ for all $i=1,2,3$, which represents the difference in items' positions in ranking $i$. 
For now, we make the simplifying assumption that there are no ties between any two items in the same ranking, so $\delta^i\neq 0$ for all $i$. (We relax this assumption later.) 
Then we have 
\begin{align}
    & \lambda^T [r^1_a, r^2_a, r^3_a] = \lambda^T [r^1_b, r^2_b, r^3_b] \\
    \Leftrightarrow 0 &= \lambda_1(r^1_a - r^1_b) + \lambda_2(r^2_a - r^2_b) + \lambda_3(r^3_a - r^3_b) \\
    &= \lambda_1 \delta^1 + \lambda_2 \delta^2 + \lambda_3 \delta^3 \\
    &= \lambda_1 \delta^1 + \lambda_2 \delta^2 + (1-\lambda_1-\lambda_2) \delta^3 \\
    &= \lambda_1 (\delta^1-\delta^3) + \lambda_2 (\delta^2-\delta^3) + \delta^3.
\end{align}

Note that the penultimate line follows from $\lambda_1 + \lambda_2 + \lambda_3 = 1$. The final equation yields a line, whose intersection with $\Lambda$ yields the \emph{boundaries between two or more indifference regions}.
 This line containing $H(a,b)$, which we denote as $L$, is summarized by two cases:
\begin{equation}
    L := \left\{\lambda\in\R^2: 
    \begin{cases}
    \lambda_2 = \frac{-\delta^3 - \lambda_1(\delta^1 - \delta^3)}{\delta^2-\delta^3}, \quad \text{if}\ \delta^2-\delta^3\neq 0,\\
    \lambda_1 = -\delta^3 / (\delta^1 - \delta^3), \quad \text{if}\ \delta^2-\delta^3= 0.
    \end{cases}
    \lambda_1+\lambda_2\leq 1, \lambda\geq 0 \right\}
\end{equation}
Note that this closed-form representation for $L \subset \R^2$ is prior to transforming to the equilateral triangle.

The closed-form representation for the endpoints of the line segment $H(a,b)=L\cap \Lambda$ is computed by testing the intersections between line $L$ and the inequalities defining $\Lambda$. 
For ease of demonstration, we use notation to say that a number divided by zero is $\infty$. Consider the following three points which represent intersections between $L$ and the constraints bounding $\Lambda$: 
\begin{align}
    (\lambda_1,0) &: \lambda_1 = -\delta^3/ (\delta^1-\delta^3), \\
    (0,\lambda_2) &: \lambda_2 = -\delta^3/ (\delta^2-\delta^3), \ \text{and} \\ 
    (\lambda_3,1-\lambda_3) &: \lambda_3 = -\delta^2/ (\delta^1-\delta^2). 
\end{align} 
Since we have assumed nonzero values for $\delta^i$, then $\lambda_i$ are also nonzero. 
Out of these three points, two belong to $\Lambda$ (i.e., they are finite, nonnegative, and sum to at most one) and are the endpoints of $H(a,b)$. 
For example, suppose $(\delta^1, \delta^2, \delta^3)  = (-1,2,2)$. Then $\lambda_1=\frac{2}{3}$, $\lambda_3=\frac{2}{3}$, and since $\delta^2-\delta^3= 0$, we say $\lambda_2=\infty$. So the two finite  endpoints are $(\frac{2}{3},\frac{1}{3})$ and $(\frac{1}{3}, \frac{2}{3})$.

\textbf{Proposition:} For input rankings with no ties, then every separating hyperplane has two distinct endpoints, in a form given by (3.7-3.9), with at least one positive component. (This means $H(a,b)$ never includes an extreme point of $\Lambda$.)

 

\begin{theorem}[Separating Hyperplane]\label{thm:cutting}
  Assume $J$ is a permutation of $\{1,2,3\}$ and $i, i'$ are distinct items in $\{1..n\}$ such that $\sigma_i(V^{j}) < \sigma_{i'}(V^{j})$ for exactly two elements of $J$. Let $j'\in J$ such that $\sigma_i(V^{j}) > \sigma_{i'}(V^{j})$, which we call the ``disagreeing'' input for items $(i,i')$. Then the hyperplane $H()$ separates the set of weights such that for all $
  \lambda\in \Lambda \cap H^< ()$,  $\sigma_i(v^\lambda) < \sigma_{i'}(v^\lambda)$, and for all $
  \lambda\in \Lambda \cap H^> ()$,  $\sigma_i(v^\lambda) > \sigma_{i'}(v^\lambda)$. 
\end{theorem}

\begin{figure}[h!]
    \centering
    \includegraphics[width=0.35\textwidth]{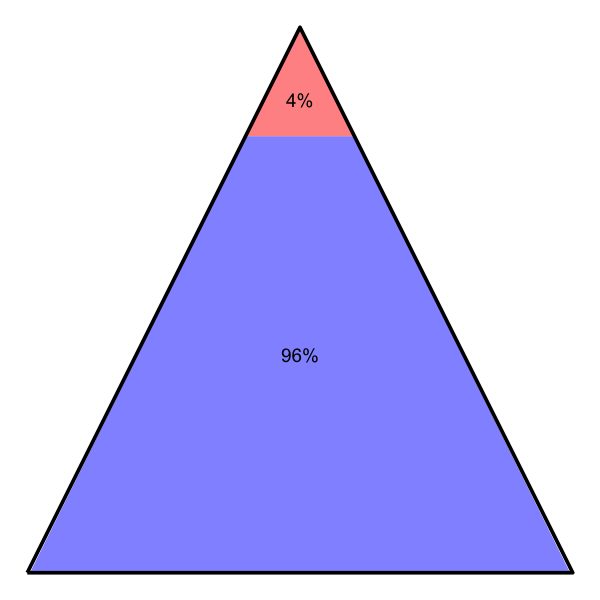}
    \includegraphics[width=0.35\textwidth]{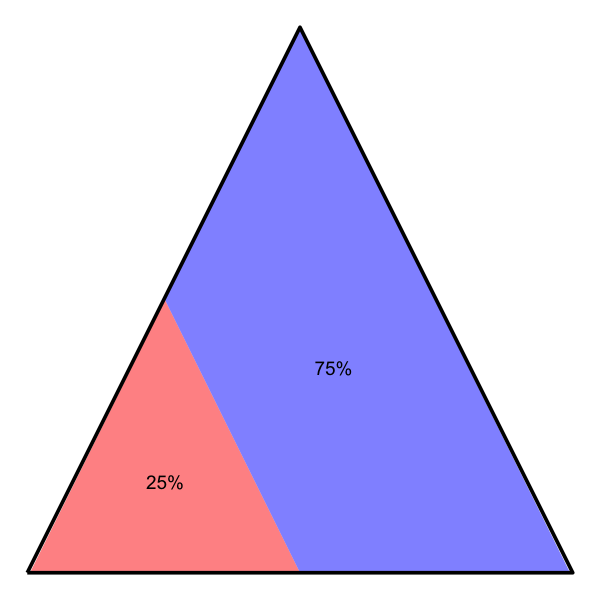}
    \caption{Pairwise item analysis. (Left) Item 1 is ranked better than 5 in 96\% of the weighted ranks. (Right) Item 2 is ranked better than 3 in 75\% of the weighted ranks.}
    \label{fig:RelativeArea}
\end{figure}

Theorem \ref{thm:cutting} provides the first analytical tool of the rank colormap. For given item pair $(a,b)$, the separating hyperplane divides the weight set $\Lambda$ into two regions, each which can be measured by area. This area is then directly interpreted as the relative proportion of weights for which $a$ is ranked better than $b$ and vice versa. Figure \ref{fig:RelativeArea} presents this item-level analysis for two item pairs. 

\subsection{Ties}
\label{subsection:ties}

When allowing ties in the input vectors, separating hyperplanes include one of the extreme points of $\Lambda$. To demonstrate in the running example, suppose input ranking vector $\b r^3$ ranks the five treatments as $[A, C, B, D/E] $ where / denotes a tie between treatments D and E in the 4th/5th rank positions. 
Then $\b r^3=[1,3,2,4,4].$
Return to equation (3.2) for the line describing the line segment containing $H(D,E)$: 

$$ 0 = \lambda_1(r^1_D - r^1_E) + \lambda_2(r^2_D - r^2_E) + \lambda_3(r^3_D - r^3_E). $$

With the tie in ranked list $r^3$, where $r^3_D = r^3_E$, $\delta^3=0$ and yields one greater degree of freedom. Then the equation of $L$ simplifies to
$$\lambda_2 = -\frac{\lambda_1(r^1_D - r^1_E)}{(r^2_D - r^2_E)}. $$

For general items $a$ and $b$, the previous representations of the line $L$ and endpoints defining $H(a,b)$ are still well-defined and correct in the presence of a tie in just one of the rankings. 


\textbf{Proposition:} For input rankings $\b r^1, \b r^2, \b r^3$ and distinct items $a$ and $b$ with exactly one tie, then the separating hyperplane has two distinct endpoints, in a form given by (3.7-3.9), where exactly one endpoint is an extreme point of $\Lambda$.

\subsection{Convexity}

\begin{theorem}[Convexity]\label{thm:convexity}
  For every ranked order $\sigma\in\Z^n$ such that nonempty $\Lambda(\sigma)$ is nonempty, $\Lambda(\sigma)$ is convex.
\end{theorem}

\begin{proof}
Proof by Induction: Let $\Lambda(\sigma)$ be a convex region. Base case: add one separating hyperplane $H$ that cuts through the region. $H$ divides $\Lambda$ into two regions, and because $H$ is linear, these two regions must also be convex. Case $n$: Assume this is true for $n$, i.e., $n$ separating hyperplanes divide $\Lambda$ into smaller regions that are convex. Case $n+1$: Add one more separating hyperplane $H$. Each of the smaller regions in $\Lambda$ is either intersected by $H$ or not. Each intersected region is then divided into two smaller regions, each of which must also be convex by the argument from the base case (i.e., because $H$ is linear).

\end{proof}

\begin{theorem}[Bound on cuts]\label{thm:bounds}
  At most ${n \choose 2}/2$ separating hyperplanes of the form $H()$ are required to fully describe the weight set decomposition for the weighted rank aggregation method.
\end{theorem}

\begin{theorem}[Neighboring Regions in Colormap]\label{thm:neighbors}
  The ranking vectors associated with neighbors sharing a border line are 1 swap away from each other. The ranking vectors associated with neighbors sharing a border point are 2 or more swaps away from each other.
\end{theorem}

\subsection{Algorithm}
\label{section:alg}

We describe two algorithms, a heuristic and an exact algorithm, for implementing the rank colormap.
A heuristic algorithm approximates the rank colormap by grid search over weight set $\Lambda$ by generating a finite number of equally spaced points in $\Lambda$, which are directly used to determine the weighted rank. A tighter grid leads to a more accurate  colormap at the cost of  more computations. 

The exact algorithm proceeds according to the following steps.  
Note that Steps 1-5 operate in the projected space before transformation to the equilateral triangle. 
\begin{enumerate}
    \item Heuristic grid search, which is primarily to collect a set of rank labels necessary for Step 3. 
    \item Compute all possible separating hyperplanes (as lines). 
    \item Compute all intersections among  hyperplanes and boundaries of $\Lambda$ (must also include the three extreme points of $\Lambda$). 
    \item Each intersection point is labeled with rank(s) (identified from Step 1) that indicate to which IRs it belongs.
    \item These intersection points form the extreme points of the IRs, which are then used to plot the convex hull, compute areas, compute centroids, etc.
    \item Transform all data structures to equilateral triangle. 
\end{enumerate}
For Step 4, a rank $\sigma$ is labeled to intersection point $\lambda\in\Lambda$ if $\b r^\lambda_{\sigma(1)} \leq \b r^\lambda_{\sigma(2)} \leq \dots \b r^\lambda_{\sigma(n)}$. 
Step 6 is further elaborated in Section~\ref{subsec:equilateral}. 

Successful completion of the exact algorithm will yield a set of convex indifference regions which cover $\Lambda$.
It is possible that the grid search of Step 1 misses one of the possible weighted ranks; however, once the convex hulls are plotted after Step 5, then this would be visibly apparent by a gap in the $\Lambda$ coverage. 
This is easily remedied by returning to Step 1 with a tighter grid search (and the spacing could be informed by the partial rank colormap). 

Advantages of the exact algorithm relate to precise boundaries and ties, which are covered in greater detail in Section \ref{subsection:ties}. An input weight at the boundary of indifference regions corresponds to more than one ranking, and this occurs when the rankings contain at least one tie.  In this case, a simple heuristic algorithm will struggle to explicitly represent a weight belonging to multiple indifference regions; it will typically  discover just one ranking at random and therefore yields boundaries that are distorted or ``fuzzy.'' An estimation of the area of an indifference region based on the heuristic algorithm will therefore be inaccurate, especially for small indifference regions. The exact algorithm computes and represents boundaries exactly, and thereby yields precise areas as well.

\begin{figure}[h!]
    \centering
    \includegraphics[width=0.95\textwidth]{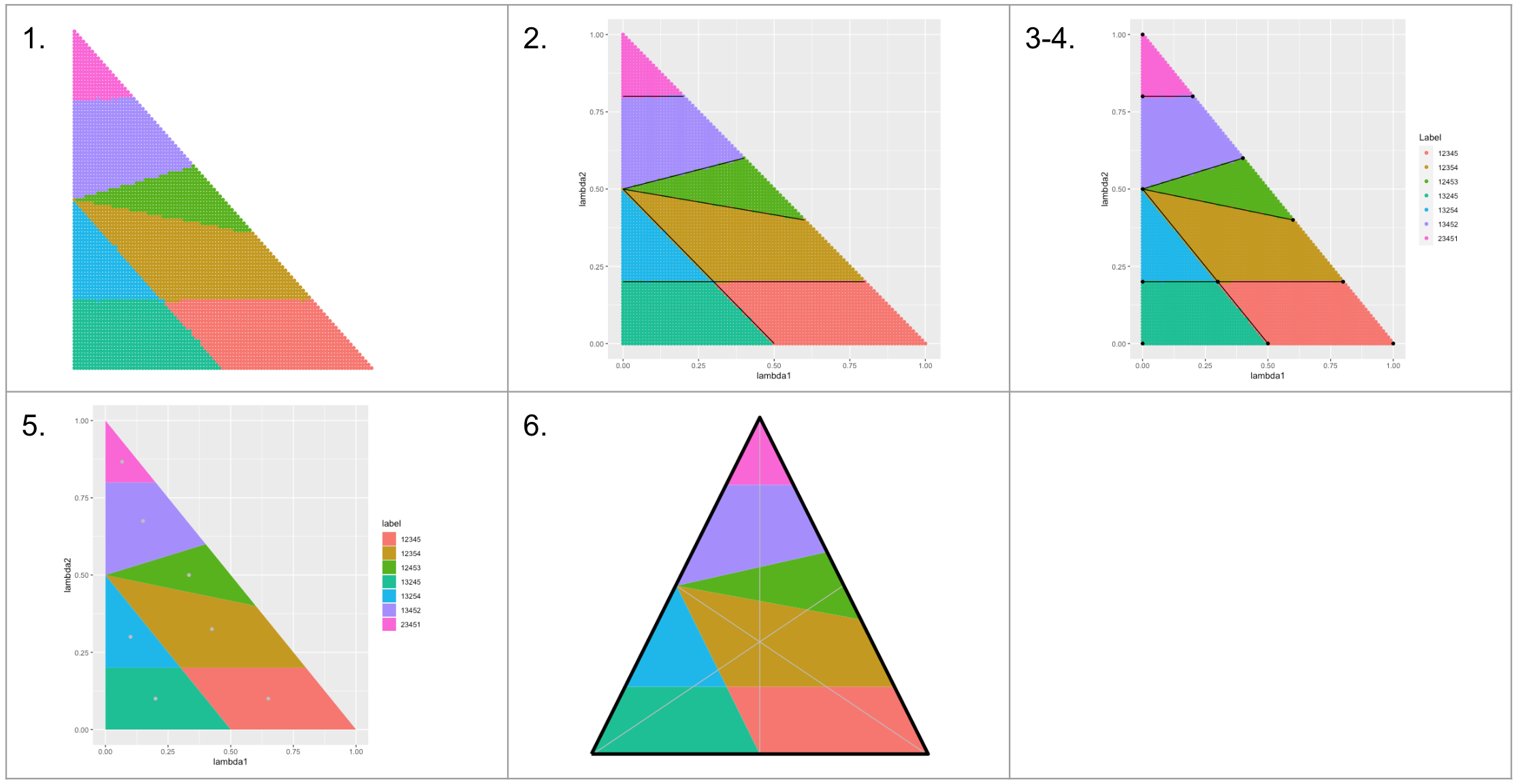}
    \caption{Illustrating the steps of the exact algorithm.}
    \label{fig:visualalgorithm}
\end{figure}

\section{Practical Notes}
\label{section:practicalnotes}

This section presents tips when implementing the ideas of the previous sections. 

\subsection{Plotting equilateral triangle}
\label{subsec:equilateral}

The easiest method for plotting $\Lambda$ in $\R^2$ is via projection by \emph{dropping} the third component, i.e., for all $\lambda\in\Lambda$, plot $(\lambda_1, \lambda_2)$. However, this results in plotting $\Lambda$ as a \emph{right} triangle. Notably, this presentation provides a biased perspective of the weight set, since the first two dimensions are plotted equivalently, but the third dimension is \emph{warped}. 
In order to transform the right triangle into the \emph{equilateral} triangle (with sides of unit length), as presented in this work, then use the following transformation:
$$\Delta (\lambda_1, \lambda_2) = (\lambda_1-0.5(1-\lambda_2), \frac{\lambda_2 \sqrt{3}}{2}). $$

\subsection{Input: rating vs ranking vectors}

\begin{figure}[h!]
\centering
\includegraphics[width=\textwidth]{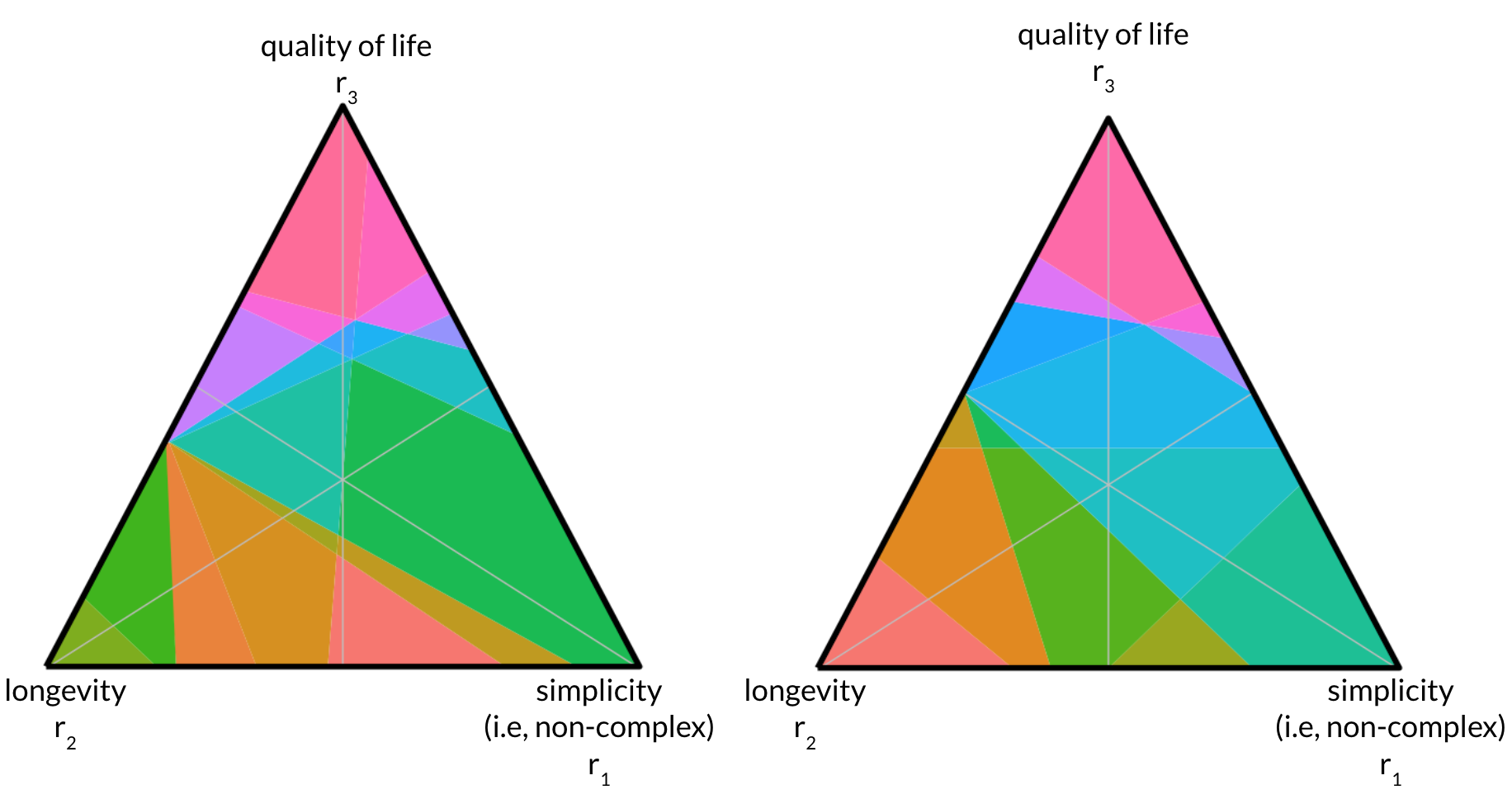} 
\caption{(left) The \emph{rating colormap} shows $|A|= 20$ regions. When the data is converted to rankings as input, the \emph{ranking colormap} (right) also has $|A|=14$ regions. Clearly, the two colormaps differ. In short, our colormap work allows for input vectors $\b r^1$, $\b r^2$, and $\b r^3$ that are either rating vectors or ranking vectors.
}
\label{fig:RatingVsRankingColormaps}
\end{figure}

In the above examples, we aggregated ranking vectors $\b r^1$, $\b r^2$, and $\b r^3$. However, many ranking and machine learning models produce \emph{rating vectors} that are then sorted to produce the ranking vectors. In fact, the ranking methods for college football, such as the Colley and Massey methods mentioned in Section \ref{section:introduction}, are ratings. The above methods can be used regardless of whether the input vectors are  rankings or ratings. Of course, ratings can be converted to rankings and then the methods of Section \ref{section:runningxCures} applied. \textbf{Or}, the ratings can remain as they are and used as input to the methods of Section \ref{section:runningxCures}. Figure \ref{fig:RatingVsRankingColormaps} shows that the colormaps can differ depending on the input. The left side of Figure \ref{fig:RatingVsRankingColormaps} uses rating vectors as input and results in $|A|=20$ regions. The right side of Figure \ref{fig:RatingVsRankingColormaps} converts these same rating vectors into rankings, which are then used as input. In this case, ranking result in $|A|=14$ regions, and the locations and areas of the regions change.

\subsection{Preprocessing: normalized rating vectors}

If $\b r^1$, $\b r^2$, and $\b r^3$ are rating vectors rather than ranking vectors, we recommend shifting and normalizing these input rating vectors so that large values in one rating do not swamp out small values in another. For example, suppose the values in the $\b r^1$ rating vector range from 10 to 200, while those in $\b r^2$ range from 0 to 1. In this case, $\b r^2$ contributes little to the aggregated ranking $\b r^a$. Related problems can occur, if, say, $\b r^3$ has negative values, ranging from $-10$ to $10$. Shift and normalize so that all three rating vectors range from 0 to 1. Shift $\b r^1$ by subtracting the positive minimum value from all entries so its new minimum value moves from 10 to 0, i.e., $\bar{\b r}^1 = \b r^1 - min(\b r^1)$. Then divide this new vector $\bar{\b r}^1$ by the new maximum value of 190, i.e.,  $\bar{\bar{\b r}}^1 = \bar{\b r}^1/max(\bar{\b r}^1)$.

\section{Further Insights}

\subsection{Item Heatmap}
\label{section:advancedcolormap}

Our tool has a feature that enables the user to focus on one item of interest. In our running example, suppose the doctor is interested in one particular treatment, Treatment $T_1$ Temozolomide,  When this item is selected in the tool's dashboard, the colormap transforms into a heatmap indicating where Treatment $T_1$ appears in each aggregated ranking in the set $A$. Lighter regions indicate that Treatment $T_1$ ranks better in the aggregated ranking associated with that region.

\begin{figure}[h!]
\centering
\includegraphics[width=0.4\textwidth]{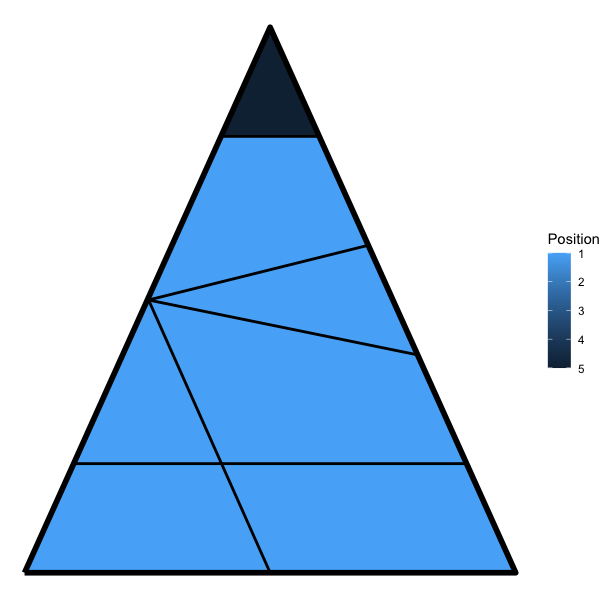}
\hfill  
\includegraphics[width=0.4\textwidth]{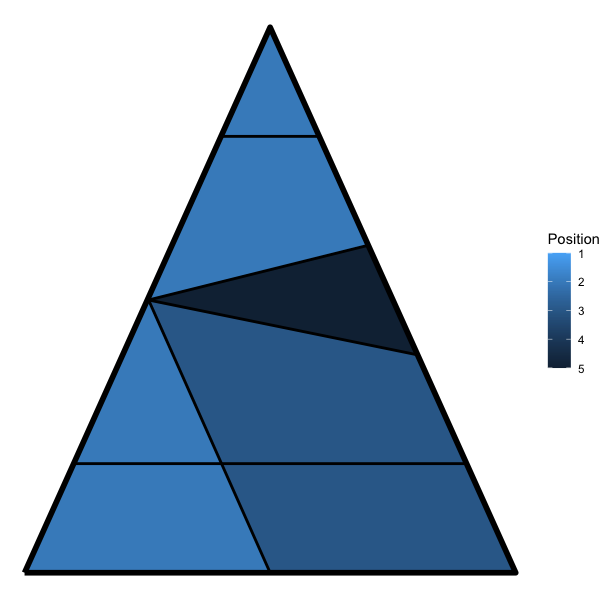}
\caption{The map on the left shows the heatmap for Treatment $T_1$ Temozolomide. Lighter regions indicate that Treatment $T_1$ ranks better in the aggregated ranking associated with that region. Treatment $T_1$ scores poorly when quality of life is the most important consideration.  The map on the right shows the heatmap for another treatment, $T_3$ Gliovac, which scores better on quality of life and not as well in the compromise area between quality of life and simplicity of the treatment regimen.
}
\label{fig:heatmap}
\end{figure}

It is clear from Figure \ref{fig:heatmap} that treatment $T_1$ Temozolomide scores worst when quality of life is the most important consideration. Contrast this with the heatmap for treatment $T_3$ Gliovac, which scores better on quality of life and not as well in the compromise area between quality of life and simplicity of the treatment regimen.

\subsection{Sensitivity Heatmap}

\begin{figure}[h!]
\centering
\includegraphics[height=6cm]{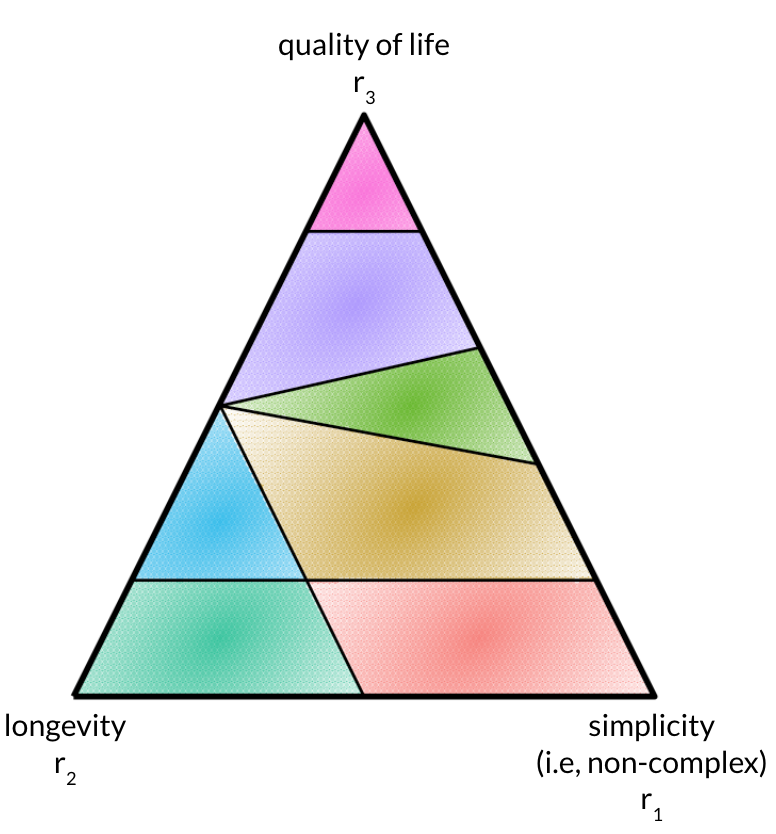} 
\caption{Sensitivity Map. The darker points near the center of a region are most robust, i.e., their ranking of treatments is least sensitive to small changes in the input weights $\lambda_i$.
}
\label{fig:sensitivitymap}
\end{figure}

Some points in an indifference region are more sensitive to small changes in the weights $\lambda_i$ than others. For example, points near an border are more sensitive than points near the centroid of a region. This can be indicated by the  darkness of the color within the region. See Figure \ref{fig:sensitivitymap}. Dark points within the region are most robust, i.e., least sensitive, to small changes in the weights $\lambda_i$. 

\subsection{Challenge for 4 or more input rankings}
\label{section:largerj}

Up to this point, we have restricted our discussion to aggregation problems that combine $j=3$ rankings (or ratings). This section determines what extends to $j \geq 4$.

\subsubsection{The $j=4$ Polytope}

\begin{figure}[h!]
\centering
\includegraphics[height=4cm]{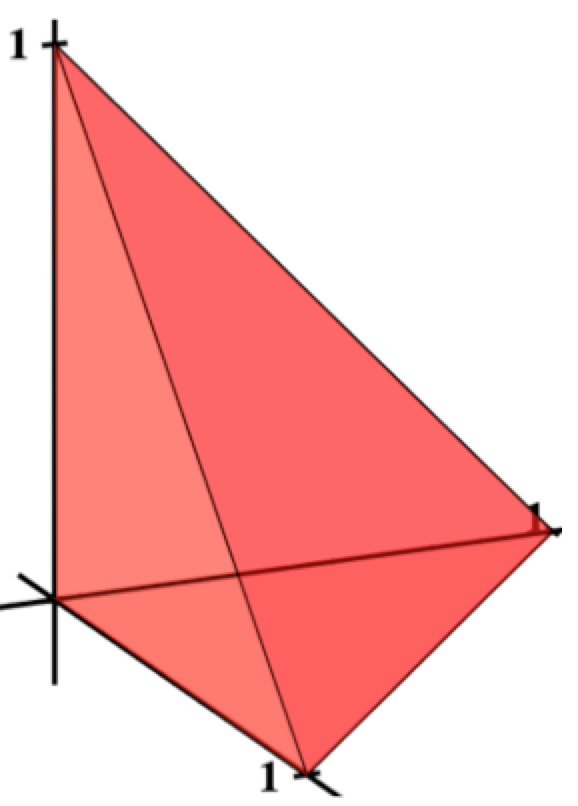} \hskip .5in
\includegraphics[height=4cm]{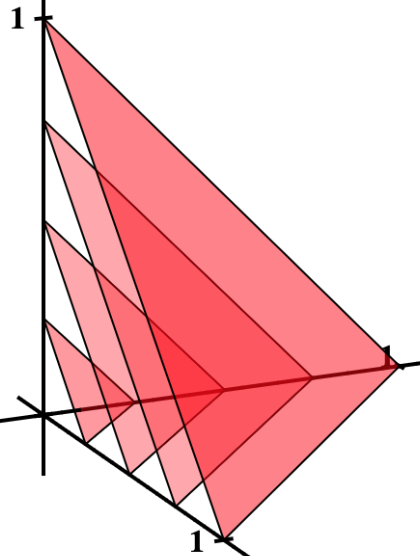}
\caption{The $j=4$ polytope in $\Re^3$ (left). Planes through the polytope for fixed values of $\lambda_4$ (right). The largest plane is the $\lambda_4=1$ plane through the polytope, which is a face of the $j=4$ polytope. The other planes are the when $\lambda_4=.75$, $\lambda_4=.5$, and $\lambda_4=.25$. The origin corresponds to $\lambda_4=0$. There are infinitely many planes at fixed $\lambda_4$ through this polytope, each with color-coded points mapped to aggregated rankings. 
}
\label{fig:planesthrupolytope}
\end{figure}

 For $j=4$, $\b r^a=\lambda_1 \b r^1 + \lambda_2 \b r^2 + \lambda_3 \b r^3 + \lambda_4 \b r^4$ and $\Lambda = \{\lambda\in\R^4_+: \lambda_1 + \lambda_2 + \lambda_3 + \lambda_4 = 1\}$ is a hyperplane in $\Re^4$. We can visualize the contribution of weights in $\Re^3$ with a polytope. Pick a point $(\lambda_1, \lambda_2, \lambda_3)$ inside or on this polytope and this forces the value of the remaining weight since $\lambda_4 = 1 - \lambda_1 - \lambda_2 - \lambda_3$. For example, the $\Re^3$ origin $(0,0,0)$ forces $\lambda_4 = 1$. The $\Re^3$ corner point $(0, 1, 0)$ corresponds to the $\Re^4$ weights $(0, 1, 0, 0)$. The  point $(\frac{1}{4}, \frac{1}{4}, \frac{1}{4})$ in the interior of this polytope corresponds to the $\Re^4$ weights $(\frac{1}{4}, \frac{1}{4}, \frac{1}{4}, \frac{1}{4})$.  See Figure \ref{fig:planesthrupolytope} (left). As a result, just as in the $j=3$ case, the $j=4$ weights can be mapped to aggregated rankings, color-coded, and plotted in and on the 3D polytope. 
 
 Each fixed value of $\lambda_4$ corresponds to a plane that slices through the polytope and hence has its own $\Lambda$ Triangle. See Figure \ref{fig:planesthrupolytope} (right).

Unfortunately, because there are infinitely many planes for the infinitely many fixed $\lambda_4$ through the $j=4$ polytope, each with color-coded points mapped to aggregated rankings, this visualization is too cluttered to be useful. However, the corresponding measures, i.e., the barchart, $A$, $|A|$, $\b X^*$, and $\b A^*$ are very useful for any size $j$.

\subsubsection{Returning to the $\Lambda$ Triangle}

This section provides an alternative approach for higher dimensional aggregation that returns to the $\Lambda$ triangle, even for $j \geq 4$. Let's consider the example application from Section \ref{section:introduction}, the U.S. News \& World Report's ranking of U.S. colleges, which aggregates 17 features. Suppose the U.S. News \& World Report is considering weight adjustments for three features, e.g., class size, first-year retention rate, and graduation rate, which are currently fixed at weights of .08, .044, and .08, respectively. This leaves the remaining features with a combined weight of .796.

By following the steps below, a rank aggregation user can \emph{see} the effects and the sensitivity of their final ranking when changes are made to the three features of primary interest. 
\begin{enumerate}
    \item The user chooses three of the $j \geq 3$ features to study and display as corners of the $\Lambda$ triangle. Without loss of generality, we label the corresponding weights $\lambda_1$, $\lambda_2$, and $\lambda_3$. For example, the U.S. News \& World Report chooses 3 of their 17 features.
    \item The remaining weights $\lambda_i$, $i=4, \ldots, j$ are fixed at their current or user-defined weights. For example, the U.S. News \& World Report may leave, or adjust, the weights of the remaining 14 features.
    \item The user chooses partition weights $p_1$ and $p_2$ for the two partitions, $\{\lambda_1, \lambda_2, \lambda_3\}$ and $\{\lambda_4, \ldots, \lambda_j\}$. For example, the U.S. News \& World Report chooses $p_1=.25$ and $p_1=.75$. 
    \item Points in the triangle $\Lambda= \{\lambda\in\R^3_+: \lambda_1 + \lambda_2 + \lambda_3 = 1\}$ are mapped to the aggregated ranking $\b r^a=p_1(\lambda_1 \b r^1 + \lambda_2 \b r^2 + \lambda_3 \b r^3) + p_2(\lambda_4 \b r^4 + \ldots + \lambda_4 \b r^4)$.
\end{enumerate}
In this way, the figures and measures from the $j=3$ sections above can be displayed for $j \geq 4$ problems.

\section{Advanced Uses and Future Work}

\subsection{Incomplete Lists}
\label{section:incompletelists}

In practice, it often happens that the ranking (or rating) vectors $\b r^1$, $\b r^2$, and $\b r^3$ are top-$k$ lists, not full lists of items. This means $\b r^1$, $\b r^2$, and $\b r^3$ are \emph{incomplete lists} as shown below. 

$$
\b r^1 = \bordermatrix{& \hbox{efficacy} \cr
1^{st} & T_1 \cr
 2^{nd}& T_2 \cr 
 3^{rd} & T_3\cr 
 4^{th}  & T_4 \cr 
 5^{th}   & T_5 \cr},
 \b r^2 = \bordermatrix{& \hbox{safety} \cr
1^{st} & T_1 \cr
 2^{nd}& T_2 \cr 
 3^{rd} & T_{10}\cr 
 4^{th}  & T_4 \cr 
 5^{th}   & T_8 \cr},
 \b r^3 = \bordermatrix{& \hbox{cost} \cr
1^{st} & T_2 \cr
 2^{nd}& T_3 \cr 
 3^{rd} & T_6\cr 
 4^{th}  & T_5 \cr 
 5^{th}   & T_4 \cr}.
 $$
 There are 8 distinct items (treatments $T_1, \ldots, T_5$, $T_6$, $T_8$, and $T_{10}$ among these three top-5 ranked lists.  Therefore, the problem is actually an $n=8$ aggregation problem with the ranking vectors below indicating the ties at the end of the rankings.
 $$
\b r^1 = \bordermatrix{& \hbox{efficacy} \cr
1^{st} & T_1 \cr
 2^{nd}& T_2 \cr 
 3^{rd} & T_3\cr 
 4^{th}  & T_4 \cr 
 5^{th}   & T_5 \cr
 6^{th}-8^{th}   & T_6/T_8/T_{10} \cr},
 \b r^2 = \bordermatrix{& \hbox{safety} \cr
 & T_1 \cr
 & T_2 \cr 
 & T_{10}\cr 
 & T_4 \cr 
 & T_8 \cr
 & T_3/T_5/T_{6} \cr},
 \b r^3 = \bordermatrix{& \hbox{cost} \cr
 & T_2 \cr
 & T_3 \cr 
 & T_6\cr 
 & T_5 \cr 
 & T_4 \cr
 & T_1/T_8/T_{10} \cr}.
 $$
When these rankings are converted to ratings by the tool, all items tied in last place receive a rating of 6, as described in the previous subsection.

\subsection{Nonlinear Utility Function}

\begin{figure}[h!]
\centering
\includegraphics[width=0.35\textwidth]{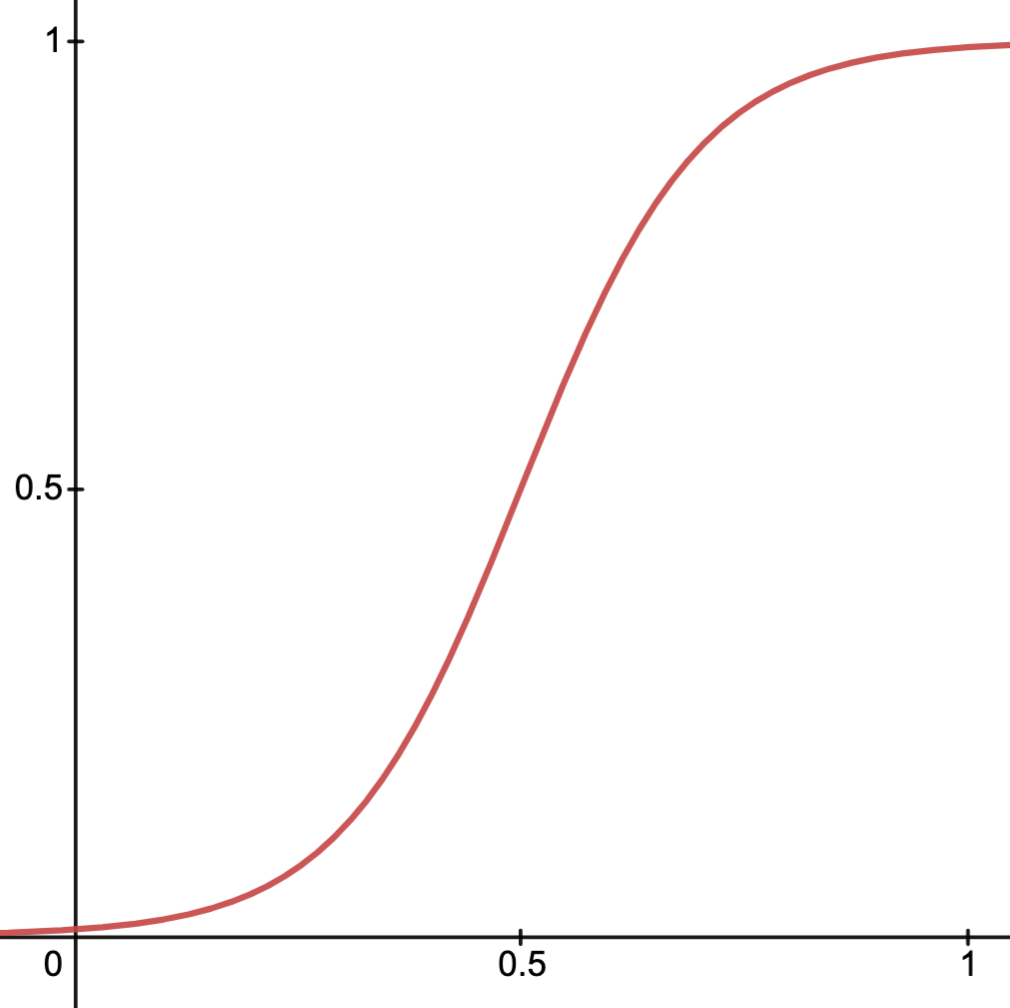} \hfill 
\includegraphics[width=0.45\textwidth]{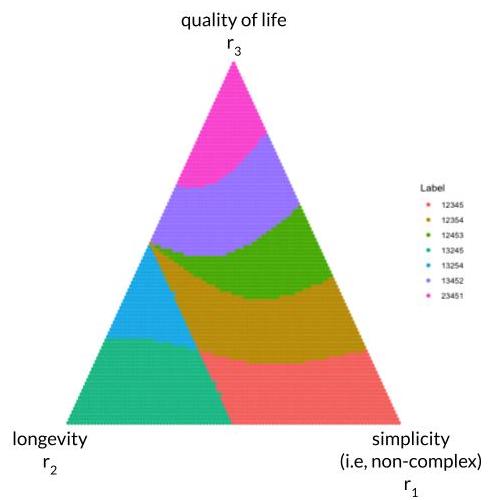}
\caption{(Left) Nonlinear function for weight $\lambda_3$ associated with \emph{cost}. 
Cost of a treatment has little impact for small values of $\lambda_3$ but then increases rapidly. 
(Right) This nonlinear transformation affects the geometry of the IRs. 
}
\label{fig:nonlinearcost}
\end{figure}

Up to this point, we have assumed that utility function for aggregating the rankings are linear in the weights $\lambda_i$. In this section, we relax that assumption. Suppose, for example, that one of the three criteria is \emph{cost}. Patients tend to say cost is unimportant, yet what they really mean is that cost matters little up to some point, then it becomes the overriding concern. In which case, a nonlinear utility function such as the one shown in Figure \ref{fig:nonlinearcost} (left) is more accurate than a linear utility function. Figure \ref{fig:nonlinearcost} (left) is a sigmoidal graph of $f(\lambda_3)= \frac{1}{1+e^{5-10\lambda_3}}$, which intuitively represents that the cost of a treatment has little impact for small values of the weight but then increases rapidly. The nonlinearity of cost can impact a doctor's plan for the treatment path (i.e., the treatment path of $T_1 \rightarrow T_4 \rightarrow T_5$ versus  $T_1 \rightarrow T_3 \rightarrow T_2$, where $T_3$ and $T_2$ are less expensive treatments). 

Assume the input rankings $\b r^1, \b r^2, \b r^3$ remain fixed, and weight $\lambda$ is still chosen from $\Lambda$. The weighted rank aggregation $C(\lambda,\b r^1, \b r^2, \b r^3)= \sum_{i=1}^3 \lambda_i \b r^i$ is now replaced with $N(\lambda,\b r^1, \b r^2, \b r^3)= \lambda_1 \b r^1 + \lambda_2 \b r^2 + f(\lambda_3) \b r^3$, where $f(\cdot)$ is positive and nonlinear.   
Given $\lambda\in\Lambda$, compute $\bar{\lambda}=(\frac{\lambda_1}{\lambda_1+\lambda_2+f(\lambda_3)}, \frac{\lambda_2}{\lambda_1+\lambda_2+f(\lambda_3)}, \frac{f(\lambda_3)}{\lambda_1+\lambda_2+f(\lambda_3)})$. Note that $\bar{\lambda}$ is  normalized so that $\bar{\lambda}\in\Lambda$. 
Furthermore, 
$$ \frac{N(\lambda,r^1, r^2, r^3)}{\lambda_1+\lambda_2+f(\lambda_3)} =  C(\bar{\lambda},r^1, r^2, r^3) .$$ 
Note that rank aggregation is \emph{scale-invariant} for positive scalars, i.e., the ranked order of values $\{v_1, \dots, v_n\}$ is the same as the ranked order of values \\ $\{\lambda v_1, \dots, \lambda v_n\}$ for $\lambda>0$. 
Hence the ranked order of aggregation by $N(\lambda,r^1, r^2, r^3)$ is identical to the ranked order of aggregation by $C(\bar{\lambda},r^1, r^2, r^3)$. 

Nonetheless, nonlinear $f(\cdot)$ leads to a weight set decomposition which may not be described by linear cutting planes as in Theorem~\ref{thm:cutting}. Figure~\ref{fig:nonlinearcost} (right) presents the resulting rank colormap, as computed by the heuristic (grid-search) method. Notably, the intersections between neighboring IRs is curved, and therefore the exact method presented in Section~\ref{section:alg} is insufficient. 

\section{Conclusions}

We invite the ranking community to use and tailor the methods of this paper and our code to their specific problems.  Ranking engineers can use it to examine the sensitivity of their ranking models, to debug code, and to derive new connections. Less technical users can use algorithms in this paper in an exploratory fashion, discovering, e.g., a treatment program that progresses through a path of treatments.

\appendix

\section*{Acknowledgments}
We thank Asher Wasserman, Lead Data Scientist at xCures, for a comment, which led to the heatmaps of Section \ref{section:advancedcolormap}.

\bibliographystyle{siamplain}
\bibliography{references}
\end{document}